\documentstyle[12pt, epsf]{article}
\begin{document}
\topmargin 0pt
\oddsidemargin 7mm
\headheight 0pt
\topskip 0mm
\addtolength{\baselineskip}{0.40\baselineskip}

\begin{center}
\vspace{0.5cm} {\large \bf Wormhole as the end state of
                    two-dimensional black hole evaporation}
\end{center}
\vspace{1cm}

\begin{center}
Sung-Won Kim$^{*}$\\
{\it Department of Science Education, \\
Ewha Women's University, Seoul 120-750, Korea}
\end{center}

\begin{center}
Hyunjoo Lee$^{\dagger}$\\
{\it Department of Physics,\\
Ewha Women's University, Seoul 120-750, Korea}
\end{center}

\vspace{1cm}

\begin{center}
{\bf ABSTRACT}
\end{center}

We present a specific two-dimensional dilaton gravity model
in which a black hole evaporates leaving a wormhole at the end state.
As the black hole formed by infalling matter in a initially static spacetime
evaporates by emitting Hawking radiation,
the black hole singularity that is initially hidden behind a timelike apparent
horizon meets the shrinking horizon.
At this intersection point, we imposed boundary conditions which require disappearance
of the black hole singularity and   generation of the exotic
matter which is the source of the wormhole as the end state of the
black hole.
These, of course, preserve energy conservation and continuity of the metric.

\vspace{2cm}
\hrule
\vspace{0.5cm}
\hspace{-0.6cm}$^{*}$ E-mail address : sungwon@mm.ewha.ac.kr\\
\hspace{-0.6cm}$^{\dagger}$ E-mail address : hyunjoo@mm.ewha.ac.kr\\
\vspace{1cm}

\newpage
\pagestyle{plain}
Hawking's discovery\cite{Hawking} that black holes radiate thermally was the beginning of
applications of quantum mechanics in black hole physics.
But this discovery has raised the information problem: according to Hawking's
calculation, when a black hole formed in collapse of a pure state evaporates,
the resulting outgoing state is approximately thermal and in particular
is a mixed state.
This conflicts with the ordinary laws of quantum mechanics
which always preserve purity. This disagreement can be avoided by introducing
the final geometry resulting from black hole evaporation to preserve purity.

In this paper we present a specific two-dimensional dilaton gravity
model\cite{BPP} in which a black hole evaporates leaving a wormhole as the end state.
Recently Hayward\cite{Hayward} proposed that black holes and wormholes are
interconvertible. In particular, if a wormhole's negative-energy generator fails or
the negative-energy source is overwhelmed by normal ordinary positive-energy matter,
it will become a black hole, and a wormhole could be constructed from a suitable
black hole by irradiating it with negative energy.

The classical two-dimensional  Callan-Giddings-Harvey-Strominger(CGHS) action\cite{CGHS} is
\begin{eqnarray}
\label{eqn:CGHS}
S_{\rm cl} &=& S_{\rm G} + S_{\rm M}       \\  \nonumber
  &=& \frac{1}{2\pi} \int d^2 x \sqrt{-g} \left( e^{-2\phi}
      \left[ R^{(2)} + 4 (\nabla \phi)^2 + 4\lambda ^2  \right]
      - \frac{1}{2} \sum ^{N} _{i=1} (\nabla f_i )^2 \right),
\end{eqnarray}
where $\phi$ is a dilaton field, $R^{(2)}$ is the 2D Ricci scalar,
$\lambda$ is a positive constant, $\nabla$ is the covariant
derivative, and the $f_i$ are $N$ matter (massless scalar) fields.
This action admits vacuum solutions, static black hole solutions,
dynamical solutions describing the formation of a black hole by
collapsing matter fields and wormhole solutions as we will see later.

To see one-loop quantum effects and back reaction, one can use
the trace anomaly, $<T^\mu  _{~~\mu} > = (\hbar/24) R^{(2)}$,
for massless scalar fields in two dimensions and  the Polyakov-Liouville action,
\begin{equation}
S_{\rm PL} = -\frac{\hbar}{96\pi} \int d^2 x \sqrt{-g(x)}
          \int d^2 x' \sqrt{-g'(x')}  R^{(2)} (x) G(x,x') R^{(2)} (x'),
\end{equation}
for which
\begin{equation}
<T^{\mu\nu} > = - \frac{2\pi}{\sqrt-g} \frac{\delta}{\delta g^{\mu\nu}} S_{\rm
PL},
\end{equation}
where  $G(x,x')$ is a Green function for $\nabla ^2$.
Here we take the large-$N$ limit, in which $\hbar$ goes to zero
while $N\hbar$ is held fixed. In that limit the quantum corrections for
the gravitational and dilaton fields are negligible, and one needs to take
into account only the quantum corrections for the matter fields.
The one-loop effective action is then $S_{(1)} = S_{\rm cl} + NS_{\rm PL}$.
In oder to find analytic solutions including semiclassical corrections,
one can modify the action as in Refs. \cite{RST}-\cite{BC}.
We use the model modified from the original CGHS model
by Bose, Parker and Peleg\cite{BPP}.

They added to the classical action (\ref{eqn:CGHS}) a local covariant term of one-loop order:
\begin{equation}
S_{\rm corr} = \frac{N\hbar}{24\pi} \int d^2 x \sqrt{-g}
           [(\nabla \phi )^2 - \phi R^{(2)}].
\end{equation}
Now the total modified action including the one-loop Polyakov-Louville term is
\begin{equation}
S = S_{\rm cl} + S_{\rm corr} + NS_{\rm PL}.
\end{equation}
We use the (extended) null coordinates $x^{\pm} = x^0 \pm x^1$ (the both coordinates
$x^+$ and $x^-$ cover the entire range($-\infty,+\infty$)) and the conformal gauge
$g_{++} = g_{--} = 0,$ $g_{+-} = - \frac{1}{2} e^{2\beta}$.
We can choose
$\phi(x^+,x^-) = \beta (x^+,x^-)$ in analyzing the equations of motion.
In the conformal gauge the equations of motion derived from $S$  are
the same as the classical ones
\begin{eqnarray}
\label{eqn:motion}
\partial_{x^+} \partial_{x^-} (e^{-2\beta(x^+,x^-)} ) &=&
\partial_{x^+}\partial_{x^-} (e^{-2\phi(x^+,x^-)} )    = - \lambda^2,  \nonumber \\
\partial_{x^+} \partial_{x^-} f_i(x^+,x^-) &=& 0,
\end{eqnarray}
while the constraints are modified by nonlocal terms $t_{\pm}(x^\pm)$ arising from
the Polyakov-Liouville action. In conformal gauge, one can use the trace anomaly
of $N$ massless scalar fields $f_i$ to obtain $<T^f _{+-}> =
-\kappa \partial_{x^+} \partial_{x^-} \beta, $ where $\kappa = N\hbar/12$, and
integrate the equation $\nabla ^\mu <T^f _{\mu\nu}> = 0$ to get the
quantum corrections to the energy-momentum tensor of the $f_i$ matter fields:
\begin{equation}
\label{eqn:tensor}
<T^f _{\pm\pm}> = \kappa[\partial^2 _{x^{\pm}} \beta - (\partial_{x^{\pm}} \beta)^2 -
t_\pm(x^\pm)],
\end{equation}
where $t_\pm(x^\pm)$ are integration functions determined by boundary
conditions.
And the modified constraints $\frac{\delta S}{\delta g^{\pm\pm}} = 0 $ become
\begin{equation}
\label{eqn:constraints}
-\partial^2 _{x^{\pm}} (e^{-2\phi(x^+,x^-)}) - (T^f _{\pm\pm} )_{\rm cl} + \kappa
t_{\pm} (x^\pm) = 0,
\end{equation}
where $(T^f _{\pm\pm} )_{\rm cl} = \frac{1}{2} \sum^N _{i=1} (\partial_{x^{\pm}}f_i)^2$
is the classical contribution to the energy-momentum tensor of the $f_i$
matter fields.

For a given classical matter distribution and a given $t_\pm (x^\pm)$ one finds
the solution for the equations of motion (\ref{eqn:motion})  with constraints (\ref{eqn:constraints}),
\begin{eqnarray}
e^{-2\phi} &=& e^{-2\beta} = -\lambda^2 x^+ x^- - \int^{x^+} dx^+ _2
     \int^{x^+ _2} dx^+ _1 [(T^f _{++} )_{\rm cl} - \kappa t_+ (x_1 ^+ )]  \nonumber \\
            & & - \int^{x^-} dx^- _2 \int^{x^- _2} dx^- _1 [(T^f_{--})_{\rm cl}
              - \kappa t_- (x_1 ^- )] +a_+ x^+ + a_- x^- + b,
\end{eqnarray}
where $a_\pm$ and $b$ are constants.
In the choice  $(T_{\mu\nu} ^f )_{\rm cl} = 0 $ and $t_\pm (x^\pm) = a_\pm = b=0$,
it means the linear dilaton flat spacetime solution.

It also has static black hole solutions corresponding to choice
$(T_{\mu\nu} ^f )_{\rm cl} = 0 $, $t_\pm (x^\pm) = a_\pm = 0$ and $b =
M/\lambda$.
In order to find the solution corresponding to the Minkowski
vacuum asymptotically, one can use Eq.(\ref{eqn:tensor}) to find the solution for which $<T^f
_{\pm\pm} ( \sigma^\pm)> = 0$ in flat coordinates $\sigma^\pm$.
The functions $t_\pm (x^\pm)$ are determined by imposing appropriate boundary
conditions that the metric is flat, such that $\beta$ and its derivatives
vanish in the asymptotically flat coordinates $\sigma ^\pm$.
Then we get
\begin{equation}
\label{eqn:vacuum}
<T^f _{\pm\pm}(\sigma^\pm)> |_{\rm boundary} =
-\kappa t_\pm(\sigma^\pm) =0,
\end{equation}
and
\begin{equation}
t_\pm(x^\pm) = \left( \frac{\partial \sigma^\pm}{\partial x^\pm} \right)
t_\pm (\sigma ^\pm) - \frac{1}{2} D^S _{x ^\pm} [\sigma ^\pm] =
\frac{1}{(2x^\pm)^2},
\end{equation}
where $D^S _y [z]$ is the Schwarzian derivative $ D^S _y [z] = \partial^3 _y
z/(\partial _y z) - \frac{3}{2} (\partial^2 _y z / \partial_y z)^2$ and we use
the fact that the Minkowski vacuum corresponds to Eq.(\ref{eqn:vacuum}).
Thus we find that the asymptotically Minkowski-vacuum solution is
\begin{equation}
\label{eqn:s-bhsol}
e^{-2\phi} = e^{-2\beta} = -\lambda^2 x^+ x^- -\frac{\kappa}{4} \ln |\lambda^2
x^+ x^- | + C,
\end{equation}
where $C$ is a constant.
This solution has two apparent horizons in the extended coordinates.

Next we turn to the dynamical scenario in which the spacetime is initially
described by one of the static solutions in Eq.(\ref{eqn:s-bhsol}), and in which the black
holes are
formed by collapsing matter fields, particularly the simple shock wave of
infalling matter described by $(T^f _{++})_{\rm cl} = (M/\lambda x^+ _0 ) ~\delta
(x^+ \pm x^+ _0 ) $, where $x^+ _0 > 0$, and $(T^f _{--} )_{\rm cl} = 0$. Then we find the solution
\begin{equation}
e^{-2\phi} = e^{-2\beta} = -\lambda^2 x^+ x^- -\frac{\kappa}{4} \ln |\lambda^2
x^+ x^- | - \frac{M}{\lambda x^+ _0 } (x^+ - x^+ _0 ) \Theta (x^+ \pm x^+ _0 )+ C,
\end{equation}
where
\begin{eqnarray}
\Theta (x^+ \pm x^+ _0) &=& 0 ~~~~~  -x^+ _0 \leq x^+  \leq x^+
_0 \nonumber \\
                        &=& 1 ~~~~~{\rm elsewhere}.
\end{eqnarray}
For all values of $M$ and $C$, the solution after shock wave is
\begin{equation}
\label{eqn:d-bhsol}
e^{-2\phi} = e^{-2\beta} = -\lambda^2 x^+ (x^- + \Delta ) -\frac{\kappa}{4} \ln
|\lambda^2 x^+ x^- | + \frac{M}{\lambda} + C,
\end{equation}
where $\Delta = M/(\lambda ^3 x^+ _0 )$.
The black hole singularity curve is
\begin{equation}
-\lambda^2 x^+ _s (x^- _s + \Delta ) -\frac{\kappa}{4} \ln
|\lambda^2 x^+ _s x^- _s | + \frac{M}{\lambda} + C = 0,
\end{equation}
and an apparent horizon, defined by $\partial_{x^+} e^{-2\phi} = 0$\cite{RST}, is
\begin{equation}
-\lambda^2 x^+ _h (x^- _h + \Delta ) = \frac{\kappa}{4}.
\end{equation}
When the apparent horizon is formed, the black hole starts radiating.
At future null infinity $ \cal I ^+$ one can calculate Hawking radiation in the
asymptotically flat coordinates  $\hat{\sigma} ^\pm $, defined by
$\lambda \hat{\sigma}^+ = \ln |\lambda x^+ |$ and $-\lambda \hat{\sigma}^- =
\ln |\lambda (x^- + \Delta ) |$,
\begin{equation}
<T^f _{--} (\hat{\sigma}^\pm)>|_{\cal I_+} = \frac{\kappa \lambda^2}{4}
\left[1- \frac{1}{(1+\lambda \Delta e^{\lambda
\hat{\sigma}^-})^2}\right].
\end{equation}
Initially the singularity is behind apparent horizon, but as the
black hole evaporates by emitting Hawking radiation the apparent horizon shrinks
and eventually meets the singularity at $(x^+ _{\rm int},x^- _{\rm int})$,
\begin{eqnarray}
\label{eqn:int}
x^+ _{\rm int} =  \frac{1}{\lambda^2 \Delta} \left[\pm \exp \left(
\frac{4( \pm M+\lambda C)}{\kappa \lambda} + 1\right) - \frac{\kappa}{4}\right],  \nonumber \\
x^- _{\rm int} = - \Delta \left[1 \mp \frac{\kappa}{4}  \exp \left(-
\frac{4( \pm M+\lambda C)}{\kappa \lambda} - 1 \right) \right]^{-1}  ,
\end{eqnarray}
where the upper signs do when $x^+ x^- < 0$ and the lower when $x^+ x^- > 0$.
The singularity become naked after the singularity and the apparent horizon have
merged.
The future evolution is not uniquely determined unless boundary conditions are
imposed at the naked singularity\cite{RST}.

Here we consider the solution to the future of the point $(x^+ _{\rm int},x^-
_{\rm int})$; at the intersection point the evaporating black hole matches to a
stable wormhole keeping continuous metric and energy conservation.
We find the boundary conditions that match the solution (\ref{eqn:d-bhsol}) continuously to a
static wormhole solution.

Before this matching, we present wormhole solutions in the (1+1)-dimensional
dilaton gravity.
The equations of motion from the classical action (\ref{eqn:CGHS}) are
\begin{eqnarray}
\frac{2}{\pi}e^{-2\phi} \left[\nabla_\mu \nabla _\nu \phi
  + g_{\mu \nu} \left( \left(\nabla \phi \right)^2
  - \nabla^2 \phi - \lambda ^2  \right)  \right]
  -  T_{\mu \nu} &=& 0            \\
e^{-2\phi} \left[R + 4 \lambda ^2 + 4 \nabla ^2 \phi
   - 4(\nabla \phi)^2  \right]  &=& 0,
\end{eqnarray}
where the first equation is derived from variation of the metric and
the second is the dilaton equation of motion.

To find the traversable wormhole solution of this theory we first
introduce the spacetime metric\cite{MT}
\begin{equation}
\label{eqn:metric}
ds^2 = - e^{2\Phi (r)} dt^2 +
  \left(1 - \frac{b(r)}{r}\right)^{-1} dr^2,
\end{equation}
where $\Phi (r)$ is redshift function which determines the structure of
the wormhole and defined by
\begin{equation}
g_{tt} = - e^{2 \Phi},
\end{equation}
and $b(r)$ is the shape function that relates to the proper radial distance
$l(r) $ from wormhole throat  by
\begin{equation}
dl / dr = \pm \left( 1-b/r  \right)^{-1/2}.
\end{equation}
In order for the spatial geometry to tend to an appropriate asymptotically flat
limit,  $\lim_{r\rightarrow \infty} b(r)$ must be finite. By
comparison with the Schwarzschild metric this implies that the mass of the
wormhole, as seen from spatial infinity, is given by $\lim_{r\rightarrow \infty}
b(r) = 2GM_{\rm w}$, where $M_{\rm w}$ is ADM mass of the wormhole\cite{Visser}.

To produce a traversable wormhole several general constraints on $b(r)$
and $\Phi (r)$ are required. The first constraint is that the spatial geometry
must have wormhole shape, i.e., two flat regions and a narrow one. Thus throat
is at minimum of $r=b=b_0$,   $1- b/r \geq 0$ throughout spacetime, and $b/r
\rightarrow 0$   as $l \rightarrow \pm \infty$ (asymptotically flat regions of
two  universes)  so $r \cong |l|$.  This constraint means that the embedding
surface flares outward, forcing $(d^2r / dz^2) > 0$ so that $r(z)$ is a minimum
at the throat in two-dimensional embedding Euclidean space with coordinates $z$
and $r$. Consequently, we have
\begin{equation}
\frac {d^2r}{dz^2} = \frac {b - b'r}{2b^2}  > 0 .
\end{equation}
Secondly, there should be no horizons or singularities since a horizon would
prevent two-way travel through the wormhole,  so $\Phi$ is
everywhere finite and $\Phi \rightarrow 0$ as $l \rightarrow \pm \infty$.
The last constraint is following: the matter and field that generate the
spacetime curvature for the wormhole must have a physically reasonable
stress-energy tensor with non-zero components $T_{\hat{t} \hat{t}} $
and $T_{\hat{r} \hat{r}} $.
In the proper reference frame of a set of observers who remain always at rest
in the coordinate system
\begin{eqnarray}
{\bf{e}} _{\hat{t}} &=& e^{-\Phi} {\bf{e}} _t  \nonumber  \\
{\bf{e}} _{\hat{r}} &=& (1-b/r)^{1/2} {\bf{e}} _r,
\end{eqnarray}
the non-zero components of the stress-energy tensor are $T_{\hat{t} \hat{t}} =
\rho =$ density of mass-energy and $T_{\hat{r} \hat{r}} = - \tau = -$ radial
tension. In this basis the metric coefficients take on the Minkowskian forms,
\begin{equation}
g_{\hat{\alpha} \hat{\beta}} = {\bf{e}} _{\hat{\alpha}} \cdot
                               {\bf{e}} _ {\hat{\beta}}
                             = \eta _{\hat{\alpha} \hat{\beta}}
                               \equiv
                               \left(
                               \begin{array}{cc} -1 & 0 \\
                                                  0 & 1
                               \end{array}
                               \right).
\end{equation}

The spacetime metric (\ref{eqn:metric}) can be written like followings by introducing
(extended) light-cone coordinates $y^{\pm} = y^{0} \pm y^1$ ( $y^{\pm}$ also
cover ($-\infty,+\infty$) ), and by choosing conformal gauge $g_{\mu \nu} = e^{2
\beta} \eta_{\mu \nu}$,
\begin{eqnarray}
ds^2 &=& -e^{2\Phi} [ dt^2 -d{r^{*}}^2]  \nonumber \\
     &=& -e^{2\beta} dy^+ dy^-
\end{eqnarray}
where $r^{*} = \int e^{-\Phi} \left( 1 - \frac{b}{r} \right)^{-\frac{1}{2}} dr$.
As $r$ grows very large, $r^* \rightarrow  \infty$.
On the other hand, when $ r $ approaches $ b = b_0$ then $r^* \rightarrow 0 $ since $ \Phi
$ is everywhere finite and we can set $ dr^* \rightarrow \infty $ at $r^* = 0$.
We then have the nonzero components of energy-momentum tensor
in (extended) light-cone coordinates,
\begin{eqnarray}
T_{+ -} &=& \frac{\partial t}{\partial y^+} \frac{\partial t}{\partial y^-}
            T_{tt} + \frac{\partial r}{\partial y^+}
            \frac{\partial r}{\partial y^-} T_{rr} \nonumber \\
        &=& e^{2\beta} (\rho +\tau)   \nonumber   \\
T_{++} &=& \frac{\partial t}{\partial y^+} \frac{\partial t}{\partial y^+}
            T_{tt} + \frac{\partial r}{\partial y^+}
            \frac{\partial r}{\partial y^+} T_{rr} \\
       &=& -e^{2\beta} \frac{y^-}{y^+}(\rho -\tau)  \nonumber   \\
T_{--} &=& \frac{\partial t}{\partial y^-} \frac{\partial t}{\partial y^-}
            T_{tt}+ \frac{\partial r}{\partial y^-}
            \frac{\partial r}{\partial y^-} T_{rr} \nonumber \\
        &=& -e^{2\beta} \frac{y^+}{y^-}(\rho -\tau). \nonumber
\end{eqnarray}
And the conservation law of energy-momentum tensor,
$\nabla^\mu T_{\mu \nu} = 0$, gives
\begin{equation}
\label{eqn:conservation}
\rho e^{2\beta} =  \frac{c}{|\lambda^2 y^+ y^-|},
\end{equation}
where $c$ is a constant.
In conformally invariant case ($T_{+-}=0$),
the equations of motion become
\begin{eqnarray}
\frac{2}{\pi} e^{-2(\phi+\beta)} \left[-\partial_{y^+} \partial_{y^-} \phi
    + 2 \partial_{y^+}\phi \partial_{y^-}\phi + \frac{1}{2} \lambda^2e^{2\beta} \right]
  &=& \rho + \tau =0    \\
e^{-2\phi} \left[\partial_{y^+} \partial_{y^-} \beta -2\partial_{y^+}\partial_{y^-} \phi
    + 2\partial_{y^+}\phi \partial_{y^-}\phi + \frac{1}{2} \lambda^2 e^{2\beta}\right]
  &=& 0.
\end{eqnarray}
Since we have gauge fixed $g_{++}$ and $g_{--}$ to zero, we should impose
their equations of motion as constraints in conformally invariant
case,
\begin{eqnarray}
\frac{2}{\pi} e^{-2 \phi}
(2 \partial_{y^+} \beta \partial_{y^+} \phi - \partial_{y^+} ^2 \phi) &=&
2\rho e^{2\beta}  \frac{y^-}{y^+}     \\
\frac{2}{\pi} e^{-2 \phi}
(2 \partial_{y^-} \beta \partial_{y^-} \phi - \partial_{y^-} ^2 \phi) &=&
2\rho e^{2\beta}  \frac{y^+}{y^-}.
\end{eqnarray}

So we get the solution determined in choice of $\phi = \beta$ as
\begin{equation}
e^{-2\phi} = e^{-2\beta} = C' + C_0 \ln|\lambda^2 y^+ y^-|
                           + C_1 y^+ + C_2 y^-  - \lambda^2 y^+ y^-,
\end{equation}
where $C', C_0, C_1$ and $ C_2 $ are constants and we can choose $c$
in Eq.(\ref{eqn:conservation}) so that
$C_0= \frac{2\pi c}{\lambda^2}=-\frac{\kappa}{4}$.
For a static wormhole the constants $C_1$ and  $C_2$ should be zero,
\begin{equation}
\label{eqn:whsol}
e^{-2\phi} = e^{-2\beta} = C' - \frac{\kappa}{4} \ln|\lambda^2 y^+ y^-|
                           - \lambda^2 y^+ y^- .
\end{equation}
Because the wormhole has forbidden range ($r<b_0$)
we have to restrict the ranges of $ y^+ $ and $ y^- $ such as $|\lambda^2 y^+ y^-| \geq 1$.
In the regions of $-\lambda^2 y^+ y^- \geq 1$,
we have to fix this solution with $C' >-1$ and $\frac{\kappa}{4}<1$
since wormhole solution has no singularities and horizons.
In the other regions, $C' > 1$ is required
from the same reason.
In this wormhole solution, one can let $C'= \frac{M_{\rm w}}{\lambda}$ like the black hole case.

Since the physical meaning of the wormhole solution is better in coordinates
where the metric is asymptotically constant on ${\cal I}_R^+$,
we set that
\begin{eqnarray}
   \lambda y^+ &=& \pm e ^{ \lambda \sigma^+}  \nonumber \\
   \lambda y^- &=& \pm e ^{- \lambda \sigma^-},
\end{eqnarray}
where $\sigma^\pm = t \pm r^* $.
These constraints preserve the conformal gauge and gives
\begin{equation}
   e^{-2\beta} = C' - \frac{\kappa}{2} \lambda r^* +
   e^{2\lambda r^*} .
\end{equation}
Thus, with the proper asymptotic flatness,
\begin{eqnarray}
  e^ {-2\Phi(r^*)} = C' e^{-2\lambda r^* } - \frac{\kappa}{2}
\lambda r^* e^{-2 \lambda r^*} + 1   \\
\rho(r^*) =-\tau(r^*) = c~ [ C' e^{-2\lambda r^* } - \frac{\kappa}{2}
\lambda r^* e^{-2 \lambda r^*} + 1]
\end{eqnarray}
As $r$ grows very large, i.e., $r^* \rightarrow  \infty$, $\Phi = 0$ and $\rho =
c $.
And when $ r $ approaches $ b = b_0$  then $r^* \rightarrow 0 $ as we mentioned before,
so at the throat
\begin{eqnarray}
   \rho &=& c~(C' + 1) \equiv \rho _0   \\
   \Phi &=& -\frac{1}{2} \ln (C' + 1) ,
\end{eqnarray}
where $c<0$ so that the constant $\rho_0 $ is a large negative value, since
the exotic matter that violates the null energy condition would
be concentrated near the throat for almost traversable wormhole.
>From a classical perspective, violation of the null energy condition is
not permitted. However we know that some quantum effects lead to
measurable experimentally verified violations of the null energy condition. In
particular, Hawking evaporation  violates the area increase theorem
for classical black holes. This implies that the quantum processes underlying
the Hawking evaporation process must also induce a violation of the input
assumption (the null energy condition - the only input assumption that seems
weak in quantum violation) used in proving the classical area increase
theorem\cite{Visser}.

Now we  give the boundary conditions that match the evaporating black hole
solution (\ref{eqn:d-bhsol}) to the static wormhole solution (\ref{eqn:whsol})
continuously at the intersection points $(x^+ _{\rm int},x^- _{\rm int})$.
The amount of energy $E_{\rm rad}$ radiated by the black hole up to
the curve $x^- = x^- _{\rm int}$ is calculated as
\begin{eqnarray}
E_{\rm rad} &=& \int^{\hat\sigma ^- _{\rm int}} _{-\infty} <T^f _{--} (\hat\sigma ^-) >
d\hat\sigma ^-   \nonumber \\
        &=& M+\lambda C - \frac{\kappa \lambda}{4} \left[ \ln
\left(\frac{\kappa}{4} \right) - 1 \right] - \frac{\kappa \lambda \Delta}{4x^-
_{\rm int}}.
\end{eqnarray}
The ADM mass of the dynamical solution (\ref{eqn:d-bhsol}) (relative to the reference
solution with $C=C_0$) is $M_{\rm ADM} = M + \lambda (C-C_0) $\cite{BPP}.
Thus we see that the unradiated mass $\delta M$ remaining as $x^- \rightarrow x^- _{\rm int} $ is
\begin{eqnarray}
\delta M &=& M_{\rm ADM} - E_{\rm rad}  \nonumber \\
         &=&  \frac{\kappa \lambda}{4} \left[ \ln
\left(\frac{\kappa}{4}\right) - 1  \right] - \lambda C_0 + \frac{\kappa \lambda
\Delta}{4x^- _{\rm int}}~ .
\end{eqnarray}
>From the constraint equations (\ref{eqn:constraints}) we find the thunderpop\cite{BPP,RST}
\begin{eqnarray}
\label{eqn:pop}
[T^f _{--} ( \hat{\sigma}^- )]_{\rm cl} &=& \frac{1}{2} \sum^N _{i=1} (\partial_- f_i )^2 \\ \nonumber
                                    &=& \frac{\kappa \lambda \Delta}{4 x^- _{\rm int}}
                                        \delta( \hat{\sigma}^- -  \hat{\sigma}^- _{\rm int} ).
\end{eqnarray}
The mass remaining after the shock wave (\ref{eqn:pop}) is
$ \frac{\kappa \lambda}{4} \left[ \ln \left(\frac{\kappa}{4}\right) - 1  \right] $
when $C_0 = 0$, and would be mass $M_{\rm w}$ of the wormhole.
Thus we try to match at intersection points (\ref{eqn:int}) the solution (\ref{eqn:d-bhsol})
to one of the static wormhole solution,
\begin{equation}
\label{eqn:oursol}
e^{-2\phi} = e^{-2\beta} =  -\lambda^2 (x^+ +A)(x^-+B) -\frac{\kappa}{4} \ln[ -\lambda^2 (x^+ +A)(x^-+B)]
+ \frac{\kappa}{4}(\ln\frac{\kappa}{4} -1),
\end{equation}
where $A=0$, $B=\Delta$  and  only the regions of
$-\lambda^2 y^+ y^- = -\lambda^2 (x^+ +A)(x^-+B) \geq 0$ is considered(see Figure 1).
The solution (\ref{eqn:oursol})
satisfies the constraints for wormhole that we mentioned below Eq.(\ref{eqn:whsol}) since
$\frac{\kappa}{4}(\ln\frac{\kappa}{4} -1) >-1 $ when $\frac{\kappa}{4} < 1$, and
has no horizons and singularities in the wormhole spacetime,  $-\lambda^2 (x^+
+A)(x^-+B) \geq 1 $.

One get the scalar curvature, $ R = 8 e^{-2\beta} \partial_{x^+} \partial_{x^-} \beta$,
which is negative in the entire region of the wormhole spacetime except the asymptotically flat region.
The no-positive property of the scalar curvature is required for an appropriate shape of the wormhole.
At the throat of the wormhole,
\begin{equation}
e^{-2\beta} |_{\rm{at ~throat}} =  e^{-2\Phi} |_{\rm{at ~throat}} =
1+ \frac{\kappa}{4}(\ln\frac{\kappa}{4} -1)
\end{equation}
and the scalar curvature is also finite and negative,
\begin{equation}
R|_{\rm{at ~throat}} = \lambda^2 \kappa \frac{1-\frac{\kappa}{4} \ln\frac{\kappa}{4} - \frac{\kappa}{4}}
{1+\frac{\kappa}{4} \ln\frac{\kappa}{4} - \frac{\kappa}{4}} <0.
\end{equation}
In asymptotic region, one can get easily $\Phi = 0$ and $R = 0$.

\begin{figure}
\epsfbox{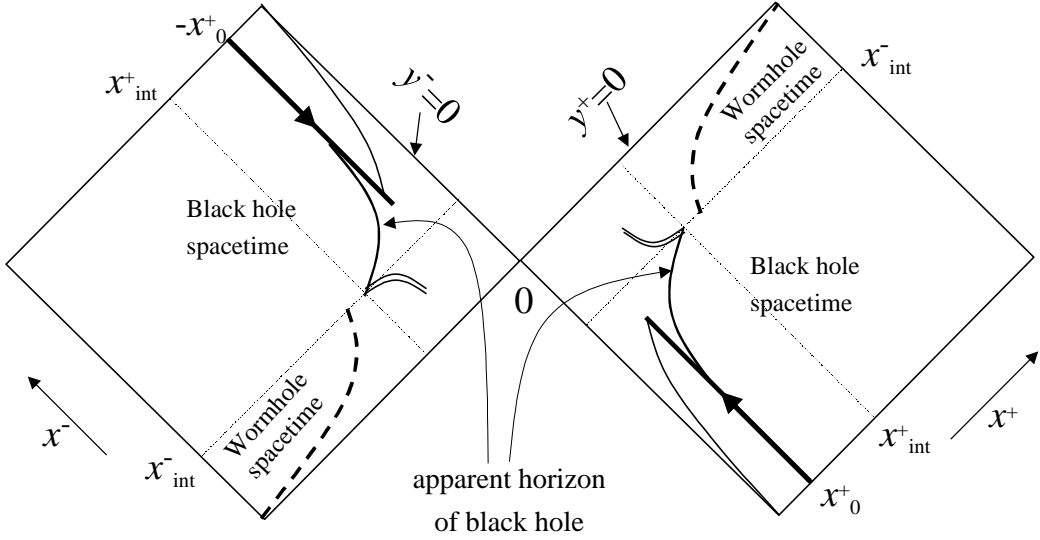}
\vspace{-10mm}
\caption{ The (extended) Penrose diagram describing the evaporating black hole
remaining the static wormhole at the end state. In this figure,
$y^+ = x^+ + A$ and $y^- = x^- + B$.
The gap between the horizon of the evaporating black hole
and the throat of the wormhole would be caused by the thunderpop at the intersection point
$(x^+ _{\rm int},x^- _{\rm int})$.
Two throat curves of the wormhole are identified
in the opposite direction.  }

\end{figure}

In this work we introduced the extended light-cone coordinates and
showed that there are  boundary conditions in which
an evaporating black hole remains a static wormhole as a candidate of end states of
a evaporating black hole.
These conditions preserve energy conservation and continuity of the metric.
This final geometry from the black hole evaporation can avoid the
information problem.

\end{document}